\documentclass[pre,twocolumn,aps,floatfix,superscriptaddress,showpacs]{revtex4}
\usepackage{longtable}
\usepackage{amsmath}
\usepackage{graphicx}
\newcommand{\beq}{\begin{equation}}
\newcommand{\eeq}{\end{equation}}
\newcommand{\be}{\begin{equation}}
\newcommand{\ee}{\end{equation}}
\newcommand{\bea}{\begin{eqnarray}}
\newcommand{\eea}{\end{eqnarray}}
\newcommand{\barr}{\begin{array}}
\newcommand{\earr}{\end{array}}

\begin{document}

\title{Kinetics of self-induced aggregation in Brownian particles.}

\author{Fabio Cecconi}
\address{INFM-SMC and Istituto dei Sistemi Complessi ISC-CNR,
         Via dei Taurini 19, I-00185 Rome Italy.}

\author{Giuseppe Gonnella}
\address{Dipartimento di Fisica, Universit\`a di Bari and \\
         Istituto Nazionale di Fisica Nucleare, Sezione
         di Bari. \\
         Via Amendola 173, 70126 Bari Italy.}

\author{Gustavo P. Saracco}
\address{Instituto de Investigaciones Fisicoqu\'{i}micas Te\'oricas y
Aplicadas (INIFTA).\\
UNLP, CONICET. Casilla de Correo 16, Sucursal 4 (1900) La Plata, Argentina.}

\begin{abstract}
We study a model of interacting random walkers that proposes 
a simple mechanism for the  emergence of cooperation in group of 
individuals. 
Each individual, represented by a Brownian particle, experiences an 
interaction produced by the local unbalance in the 
spatial distribution of the other individuals. This interaction results in 
a nonlinear velocity driving the particle trajectories in 
the direction of 
the nearest more crowded regions; the competition among different 
aggregating centers generates nontrivial dynamical regimes. 
Our simulations show that for sufficiently low randomness, 
the system evolves through a coalescence behavior characterized by 
clusters of particles growing with a power law in time. 
In addition, the typical scaling properties of the 
general theory of stochastic aggregation processes are verified.
\end{abstract}


\maketitle

\section{Introduction}
A process frequently encountered in the study of chemical and
physical phenomena is the aggregation of small particles 
joining each other to form larger spatial structures and clusters.
The comprehension of the general properties of aggregation processes
constitutes a cross-disciplinary interest for 
pure and applied research \cite{Klett,Flory,Granu,Mehlig,Turbo,Galaxy} 
with broad implications to engineering and industrial 
technology \cite{Aggreg}.

Aggregation is also a basic process in those biological systems where 
cooperation activity among individuals usually involves social 
behaviors. A well known example     
is {\em animal grouping} where an ensemble of individuals belonging to
the same species live together into organized communities such as insect 
swarms, mammal herds, fish-schools, and bird flocks \cite{Okubo,Flierl}.
The emergence of cooperation is still one of the most
puzzling mechanism occurring in biology, mainly because social and
altruistic behaviors are against evolutionary selection which, instead, 
promotes antagonism and competition. 
Mathematical population biology
\cite{MathBio,Hirsch,May}
and Game Theory \cite{Axelrod,Smith,Dilemma,Japan} have always attempted to
reconcile altruism with evolutionary selection 
trying to establish rigorous and quantitative basis to the emergence 
of cooperation among individuals.
One of the first approaches, known as 
{\em kinship theory}, has been elaborated by W.D.~Hamilton~\cite{Hamilton}. 
It is based on the principle of {\em kin selection} 
and allows altruism to arise among siblings provided 
they share enough common genes. Although the Hamilton's principle 
strictly works for related individuals, it has been also applied to cases 
of absence of kin recognition as a rather general explanation for altruism. 
Alternatively, when genetic arguments cannot be manifestly invoked because 
cooperation involves unrelated individuals,     
it is reasonable to resort to minimalist phenomenological models  
describing the mutual advantage to form a group or to make
coordinated movements \cite{Leader}
in terms of effective ``social'' interactions.    
Synchronized and collective behaviors, on the other hand, are 
the result of ``short-'' and ``long-range'' interactions between different 
units which, in social biological systems, are not very dissimilar to
those used in the description of fluids and condensed matter.
In this perspective, it is reasonable to hope that some general features of 
collective behaviors of living organisms can be simulated and 
classified through the same principles of statistical 
mechanics successfully applied to understand the structural organization of 
the matter.

General and simple theoretical descriptions of cooperation in social
communities requires two principal ingredients: i) the space of states,
where each point represents the status of the individuals and ii) the strategy, 
that is the rule according to which the individuals (players) decide to 
change their status in response to partial or complete information about 
the actions of the other players \cite{Smith,Cooperat}. 
The prevalence of cooperation is the outcome of those 
strategies
rewarding altruistic acts and punishing defections.
These rules should be given as simple and generic as possible in order to
capture the essential and universal aspects of the problem without making the
theoretical approach extremely complex or too computationally demanding.

In a previous work \cite{CeccoPRL,CeccoPRE} we studied 
a model suggested by Sigmund and Nowak \cite{SigNow} to test 
the hypothesis of 
the emergence of cooperation by {\em indirect reciprocity} 
among unrelated individuals. 
In their game theoretical approach, Sigmund and Nowak assumed
that cooperation to work in evolved social systems requires 
the knowledge of ``reputation or status'' of their members (players). 
Thus a dynamical coordinate, $S$, 
the {\em image score}, is assigned to each player signaling his/her 
reputation or status to the group.  
$S$ is updated according to the altruistic or selfish acts
made by players in the past and it is an information accessible to the 
whole community.
At each turn of the game, a randomly selected player with score $S$ 
assesses the possibility to provide help only to
those opponents with a score greater than $S$. 
The image score, thus, determines the selfish or altruistic strategy. 
Altruism, despite the cost, is preferably since it increases 
$S$ that, conversely, is lowered by a defection. In this scheme, cooperation  
occurs thanks to the principle ``help and you will be helped''.

A simpler version of this problem was recast,  
in a recent paper \cite{CeccoPRL}, in  
terms of a non-linear Fokker-Planck equation
for the population of individuals with a certain image score.
The equation had a non-local drift term that characterized the strategy.
The peers exert a sort of ``pressure'' on each other within a finite 
range of influence in order to uniform their image score to the majority.
In this sense, the model produces an aggregation mechanism in the space of
scores driven by a population gradient; aggregated states correspond to
situations where cooperation is achieved.
The model, upon changing the system parameters, exhibited a transition from
an aggregation
behavior to a uniform state with no prevalence of selected image scoring.

This model, in another context and with proper changes, 
can also be used to describe pattern formation and chemotaxis
phenomena where diffusion competes with a drift induced by
chemical or population gradients \cite{Murray,
Bacteria,Lopez,Leuko,EColi}.

In this paper we focus on the dynamics of the aggregation process which
was not considered in Ref.~\cite{CeccoPRL}. To this aim, we
employ a discrete, or individual based, variant of that model,
where a set of individuals modify their status according to
stochastic differential equations
corresponding to a Brownian motion with a drift
induced by the spatial distribution of other walkers.
This model can also be interpreted as a collection of interacting random
walks \cite{IntRW} where 
the path coalescence depends on the balance between the diffusion
and the nonlocal drift.  
The particle-based modeling permits one to track the
state and position of each component of the system. 
It is appropriate to study the fluctuations in this kind of 
aggregation dynamics characterized by the formation and merging 
of growing clusters. 
We shall see that, differently from previous proposed mechanisms,
the coalescence of trajectories does not emerge as a result
of direct interactions between particles \cite{Sire} or via 
reaction-diffusion mechanism \cite{Okubo},
but rather it is the consequence of the drift term sensible
to density fluctuations occurring even at distances relatively large.  
We found that in this model, aggregation and clustering follow the general
scaling behavior of stochastic coalescence phenomena 
\cite{Scaling,Leyvraz,Redner} 
with power law for the kinetics of the number and 
average mass of clusters. 

The paper is organized as follows. In sect. II we describe the model
and the basic features of its dynamics. Sect. III is devoted to
the presentation and discussion of numerical results. Finally, in
sect. IV, we draw conclusions.

\section{Model of Interacting Individuals}
The model consists in a system of $N$ units (individuals) which change their
state $x_i$ according to a sort of majority rule.
The variable $x_i$, referred to the $i$-th member, might indicate 
the reputation score in indirect cooperation models, 
the position in a possible chemotaxis description, or some other amplitude
characterizing the role of an individual within a population 
biology framework.
Hereafter, without loss of generality, it is convenient to adopt the
terminology of spatially distributed systems, thus $x$ will be a
spatial coordinate. The formulation, however, can apply to other and 
different contexts.

We assume that each individual changes $x_i$ according
to the non-local stochastic equation of motion
\begin{equation}
\dot{x_i} = v(x_i) + \sqrt{2D} \xi_i
\label{eq:evol1}
\end{equation}
with the drift $v$ defined by the formula
\begin{equation}
v(x_i) = \lambda \frac{w_{+}(x_i,t) - w_{-}(x_i,t)}
                      {w_{+}(x_i,t) + w_{-}(x_i,t)}
\label{eq:evol2}
\end{equation}
where $w_{\pm}$ are given by
\begin{equation}
w_{\pm}(x_i) = \sum_{j} \Theta[\pm(x_j - x_i)] \exp(-\alpha|x_j - x_i|) 
\label{eq:evol3}
\end{equation}
The velocity at which an individual ``decides'' to move to the
left or to the right is the result of his perception of the unbalance,
$w_{+} - w_{-}$, between the populations at his left and right.
This perception is simply modeled by the exponential weight with
a coefficient
$\alpha = 1/r_0$ that defines the {\em sensing distance}, {\em i.e.}
the range $r_0$ within which one individual still perceives the presence and
the influence of the other members of the group ($|x_i-x_j| \leq r_0$).
This means that, unless $\alpha=0$, only partial information about the
population is available to each individual.
In this scheme, $v_i$ is likely to point towards 
the most populated regions if not too far from the $i$-th player;
the evolution rule (\ref{eq:evol1},\ref{eq:evol2}), therefore, defines an
aggregation process where individuals preferably migrate, advected by $v$,
towards the regions with higher local population.
$\Theta(s)$ is the unitary step function.
A random noise $\xi$ of zero average and correlation
$$
\langle \xi_i(t) \xi_j(t') \rangle = \delta_{ij} \delta(t-t')
$$
incorporates some degree of uncertainty or randomness in the dynamics
of the system, introducing the possibility for the individuals to change
even randomly their state.
In a possible chemotaxis framework \cite{Murray}, the system
(\ref{eq:evol1},\ref{eq:evol2}) represents the motion of biological
organisms
that, while performing a random walk with diffusion coefficient $D$, 
secrete pheromones  
whose concentration turns to be proportional to the number of individuals 
at a given place. Then, the organisms following the 
concentration gradient feel an attraction towards the nearest most 
crowded zones \cite{KS}. 
The parameter $\lambda$ sets the range of the velocity excursion; 
the normalization factor $w_{+}(x_i) + w_{-}(x_i)$
in  Eq.~(\ref{eq:evol2}) makes the velocity bounded within 
$[-\lambda,\lambda]$ implying that individuals have a 
limited response velocity as it happens in realistic cases.
This maximal speed is reached when
$w_{+}(x_i) = 0$ or $w_{-}(x_i) = 0$, {\em i.e.}
when the whole population lies completely on the right/left with respect to
$x_i$. The $i$-th individual, then, will move ballistically
with the absolute maximal velocity $|v_i|=\lambda$ to join the other $N-1$
members independently of $N$. 
Thus the drift $v(x_i)$ is a nonextensive
parameter defining the model.
\begin{figure*}
\includegraphics[clip=true,keepaspectratio,width=16.cm]{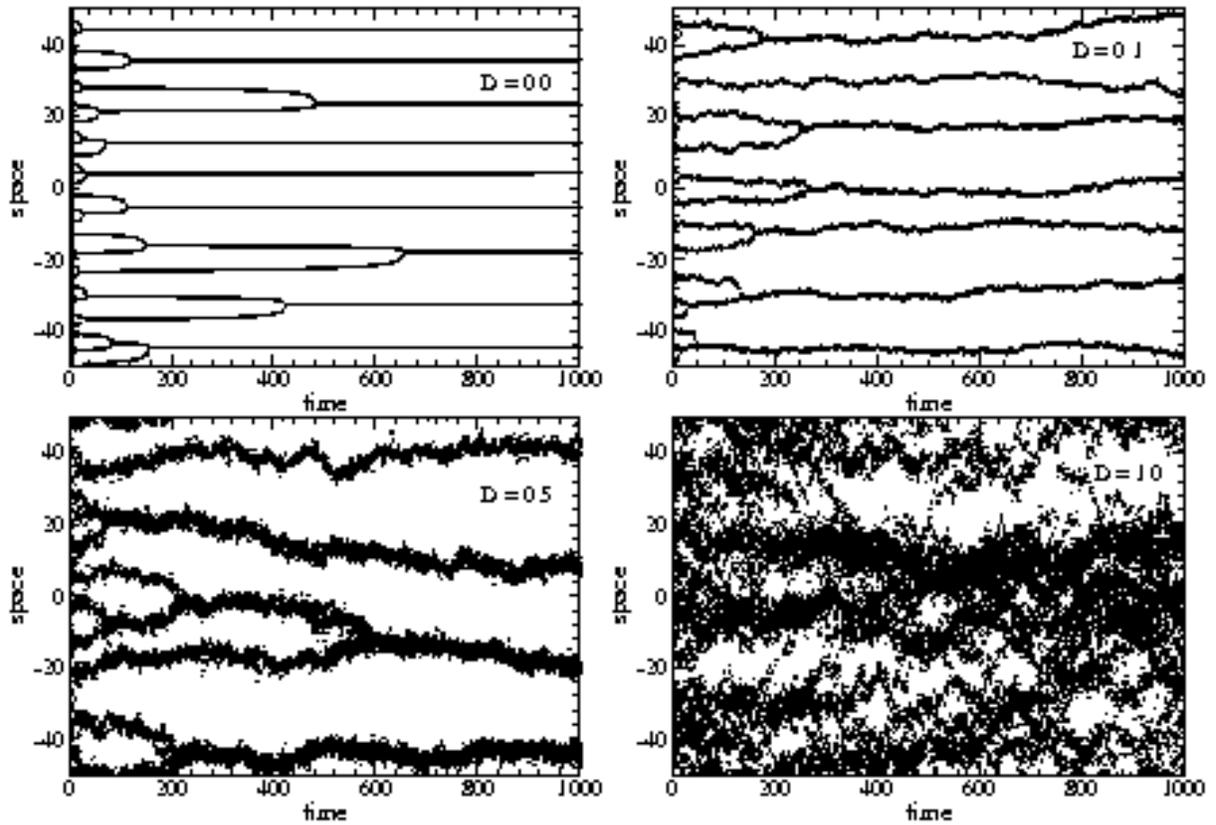}
\caption{Trajectories of $N=100$ particles obtained 
by numerical integration of Eq.~\eqref{eq:evol1} with time-step $h=0.01$,
starting from  uniformly distributed 
initial conditions in a range 
$L=[-50,50]$. The four panels, referring to   
parameters $\alpha=1$, $\lambda = 1$ and noise $D = 0$, 
$D = 0.1$, $D = 0.5$ and  $D = 1$, 
show the influence of the randomness on the system phenomenology. 
True path coalescence is observed only for $D<1$.}
\label{fig:config}
\end{figure*}
Equation~(\ref{eq:evol1}) can be cast in a
dimensionless form with the proper rescaling of time
$t\to t/(\alpha \lambda)$ and space $x \to x/\alpha$,  where only one
independent parameter survives.
In the natural units ($\alpha = \lambda = 1$), the variance of the noise
is renormalized to $D \to \alpha D/\lambda$ \cite{CeccoPRL}. Hereafter 
therefore we work with the values $\alpha = \lambda = 1$. 

In our study, we consider the effects of varying both the 
randomness and
the density of individuals in order to see how they affects the kinetics of
aggregation.
For low noise levels ($D < D_c$) the system has a strong
propensity to aggregation
and almost independently of the initial conditions the particles 
form clusters that collapse each other to generate new larger clusters. 
Figure \ref{fig:config} provides some instances of the basic
phenomenology of the system dynamics, 
for $D=0$, $D=0.1$, $D = 0.5$ and  $D = 1$. 

In small systems, the process continues until only one
cluster survives with a well
defined size  and particle distribution \cite{CeccoPRL}.
However, in larger systems $N \to \infty$ with random initial 
configuration, 
this final state is practically inaccessible to simulations
because the transient regimes become extremely long. 
In this case the aggregation kinetics is the
dominating and relevant feature of the system behavior.

\begin{figure}[h]
\includegraphics[clip=true,keepaspectratio,width=8.0cm]{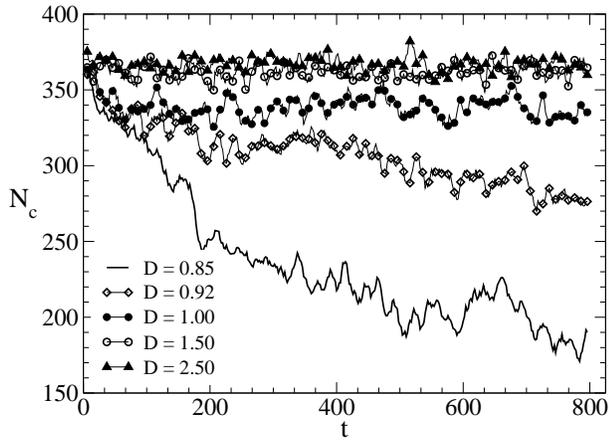}
\caption{Evolutions of the number of clusters formed in a system
with density $\rho = 10$ and
different variance $D$ of the noise, indicated in the inset.
The threshold $D_c=1$ separates aggregation
from uniform distribution of particles.}
\label{fig:transition}
\end{figure}

For $D>D_c$, the dynamics of the system loses its tendency to
cluster and particles basically perform a standard random walk inside the
one-dimensional box containing them.
This kind of non-equilibrium transition is clearly illustrated by
figure~\ref{fig:transition} where we plot the evolution
of the number of clusters $N_c$ that are present in the system. We
define a cluster as a set of particles with mutual distances
$|x_i - x_j|$  below a given cutoff $\varepsilon$  
that we shall call resolution.
We see that for noise below $D_c=1$,  $N_c$ fluctuates
but exhibits a decreasing trend in time,
signaling that clusters coalesce with each other so
their average number decreases.

\begin{figure}
\includegraphics[clip=true,keepaspectratio,width=8.cm]{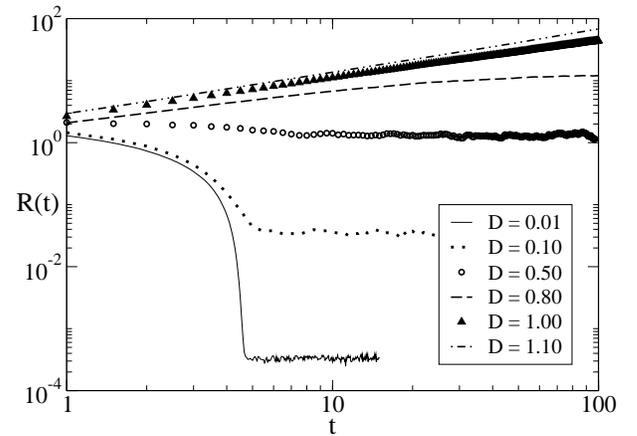}
\caption{Log-log plot of the system variance $R$ versus time in 
spreading experiments with $N = 500$ particles at different
noise level showing the crossover from localized to delocalized 
evolutions. Each curve is the result of the average over $200$ independent
runs.}
\label{fig:spread}
\end{figure}

If $D> D_c$, after a small transient, 
the number of cluster fluctuates around a stationary value.
A similar behavior is observed in {\em spreading} experiments 
where the system evolves from a set of spatially localized initial 
conditions. This kind of numerical simulations requires
no boundary conditions to let the system free to occupy a
broader and broader portion of the x-axis. 
The time evolution of the variance of the system 
with respect to its center of mass 
\begin{equation}
R(t) = \frac{1}{N^2} \sum_{i,j>i} 
\langle [x_i(t) - x_j(t)]^2 \rangle
\label{eq:spread}
\end{equation}
is shown in Fig.~\ref{fig:spread} for different values of $D$ and 
averaged over independent runs.  
Upon rising $D$ from zero, $R(t)$ displays a crossover from a regime
in which it decreases with time or remains bounded to 
a regime where spreading occurs
as $R(t)$ grows with time from the initial value $R(0) = \sigma_0^2 $. 
This basically confirms 
the scenario already described for the number of clusters:
when the dynamics is dominated by noise effects, the clusters
become unstable and evaporate.

The picture emerging from the simulations
on the existence of the bifurcation point $D_c=1$ is affected 
by finite size effects allowing only an approximate  
identification of $D_c$. However, the presence of a transition at 
the exact value $D_c=1$ can be supported by the following 
analysis~\cite{CeccoPRL}.  
It is convenient to consider 
the Fokker-Planck (FP) equation associated to 
the Langevin equation~(\ref{eq:evol1}) and observe that the latter  
can be derived formally from a local potential  
$U(x_i) = -\ln\{\phi(x_i)\}$,
\begin{equation}
\dot{x_i} = 
-\frac{\partial U(x_i)}{\partial x_i}  + \sqrt{2D} \xi_i =
\frac{1}{\phi(x_i)}\frac{\partial \phi(x_i)}{\partial {x_i}}  + 
\sqrt{2D} \xi_i 
\label{eq:langev}
\end{equation}
where 
$$
\phi(x_i) = \sum_{j \neq i} \exp(-|x_j-x_i|)
$$
and $D$ indicates the renormalized noise variance.  
In limit of large number of particles and in the mean-field like approach, 
the sum over  $j$ can be replaced by an integral 
involving the single particle density $\rho(x,t)$, 
\begin{equation}
\phi(x,t) = \int dy\, \rho(y,t) \mbox{e}^{-|y - x|}\,.
\label{eq:phi}
\end{equation}
This formula clarifies that the dynamics of each walker can be
regarded as it occurs in a time dependent mean potential generated by the 
local environment produced by the position of the other walkers. 
The integral equation~\eqref{eq:phi} can be transformed into a differential
equation for $\phi$ by a double differentiation of both members
\begin{equation}
\partial^2_x \phi(x,t) - \phi(x,t) = -2 \rho(x,t)\;.
\label{eq:Poisson}
\end{equation}
The FP equation governing the evolution of the 
density of particles obeying Eq.~\eqref{eq:langev}
is     
\begin{equation}
\partial_t \rho = -\partial_x(\rho/\phi \partial_x \phi) + 
D \partial_x^2 \rho\;.
\label{eq:FP}
\end{equation}
It admits a stationary implicit solution $\rho_s(x) = C[\phi_s(x)]^{1/D}$
with $C$ a positive integration constant determined by the normalization of 
$\rho$. The solution has to be made explicit through the substitution 
$\rho_s(x)$ into Eq.~\eqref{eq:Poisson},
$$
\phi_s'' - \phi_s = -2 C \phi_s^{1/D}
$$
that integrated after a multiplication for $\phi_s'$ of both members 
yields
\begin{equation}
\frac{1}{2}\phi_s'^2 - \frac{1}{2} \phi_s^2 + 
\frac{2DC}{1+D}\phi_s^{1+1/D} = E \;, 
\label{eq:energy}
\end{equation}
with $E$ being another integration constant. The 
normalizability of $\rho_s(x)$
and, as a consequence, of $\phi_s(x)$ for infinite systems  
requires that $\phi$ vanishes enough faster when $x\to \pm\infty$   
implying that $E=0$.    
Equation~\eqref{eq:energy}, provided that $x$ is interpreted as a time
coordinate,
corresponds to the motion of a particle with unitary mass in a 
potential  
$$
V_p(\phi_s) = - \frac{1}{2} \phi_s^2 + \frac{2DC}{1+D}\phi_s^{1+1/D}\;.
$$
The properties of $V_p$ explain the presence or absence of
coalescence. Indeed,  
when $D>1$ the motion along the orbit $E=0$ reduces to the fixed point 
$\phi_s(x)=0$ [$\rho_s(x)=0$], while nonzero and localized 
solutions $\phi_s(x)$ can be found only for $D<1$. 
This indicates the presence of a change in the system 
behaviour occurring when $D$ goes through the critical point $D_c=1$.

A linear stability analysis over the nonlinear FP equation leads to the 
same conclusion. Let $\delta \rho(x,t) = \exp(ikx-\mu t)$ a 
spatially modulated perturbation of vector $k$ over a uniform 
density background. A substitution into the integral 
\eqref{eq:phi} gives $\phi(x,t) = \exp(ikx)/(1+k^2)$ that, in turn, 
plugged in the FP equation determines 
the simple condition $\mu = k^2(D - 1)$. 
Then the perturbation $\delta \rho$ is exponentially 
reabsorbed in time for $D>1$ 
so that the uniform state is stable, whereas for $D<1$ the perturbation
is unstable signaling drastic changes in the system evolution. 

In the following we shall focus on the characterization of the aggregation
dynamics for $D<D_c$
by monitoring and quantifying the properties of the
clustering kinetics resulting from path coalescence. 

\section{Aggregation Dynamics}
We study the dynamics of the system via simulations of
Eqs.~\eqref{eq:evol1} and \eqref{eq:evol2} numerically integrated
through standard second order Runge Kutta algorithm for stochastic
equations \cite{RKstoc}. All of our
simulations, unless differently indicated, started from disordered
uniform configurations with particles contained in a one-dimensional box of
a fixed size $L$. We used a time step $h=0.01$, which is a
fair compromise between statistical accuracy of the algorithm and the
length of simulations.    
In the range of explored parameters 
we verified that a refinement of the time step 
does not affect the results. 
During the runs, when a particle close to the
boundaries escapes the box, it is reinjected on the other side.
Such reinjection events are rare because, in the noise range we
worked, the drift term~\eqref{eq:evol2} is dominant and the
particles close to the boundaries move preferentially towards the
bulk of the systems; thus border effects are really irrelevant.
However, to check that the results are not affected by the
re-injection through the boundaries, we repeated some of the
simulations with fully periodic boundary conditions ({\em i.e.}, 
considering system on  a ring of length $L$). The different
implementation of the boundary conditions did not alter the
results. 
In the periodic boundary conditions the left and right position 
with respect to an individual located at $x_i$ is specified by considering 
that each individual divides the ring in two 
halves of length $L/2$ each, accordingly,  
all individuals laying on the left or on the right of $x_i$  can be 
identified without ambiguity. 

In this section, we analyze and characterize the aggregation 
kinetics observed in the system simulations at  
low noise regimes, where the dynamics is 
dominated by the drift term. 
As shown in Fig.~\ref{fig:config}, aggregation starts by forming 
small clusters which, in turn,  
collide and merge to generate larger and larger structures. 
The larger clusters exhibits low coalescence rates and acts as 
coagulation centers by absorbing single particles or smaller clusters.

\begin{figure}
\includegraphics[clip=true,keepaspectratio,width=8.cm]{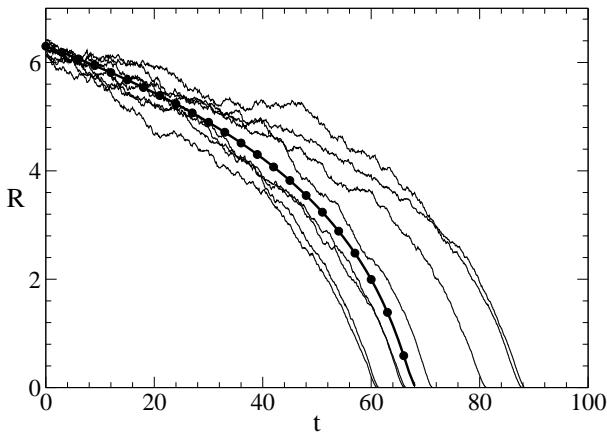}
\caption{Time behavior of $R(t)$ [Eq.~\ref{eq:spread}] for a system with
$N = 100$ particles, uniformly distributed with density $\rho=10$
and $D = 0.01$. The pointed line
indicates the corresponding quantity obtained by the numerical
solution of the differential Eq.~(\ref{eq:simple}).}
\label{fig:theory}
\end{figure}

A first quantitative attempt to describe the aggregation 
kinetics can be made by the following argument that considers the 
very ideal case of  only two clusters,
approximated as flat particle distributions in the intervals
$[X_i-\delta/2,X_i+\delta/2]$ around their centers of mass $X_i$
($i=1,2$). By assuming that they attract each other without changing
too much of their structure, the average
distance between the centers of mass $\Delta = X_2 - X_1$
contracts in time according to the equation (see the Appendix A):
\begin{equation}
\dot{\Delta} = -\frac{2}{1 + b\;\exp(\Delta)} \,
\label{eq:simple}
\end{equation}
where $b$ is a parameter depending on the cluster geometry, in the
specific case $b = 2/\{1 + \exp(\delta/2)\}$; $b$ 
can also be interpreted as a tunable parameter of a
fitting procedure. In this
ideal situation, the mean spreading, Eq.~(\ref{eq:spread}),  
verifies the relation $R(t) = \Delta^2(t)/4$.  
In Fig.~\ref{fig:theory}, we plot the theoretical spreading
$R(t)$ obtained by the integration of Eq.~(\ref{eq:simple})
along with the corresponding result extracted from simulations 
on a system with density 
$\rho = 10$, $N = 100$ particles and starting from random configurations. 
We observe the
formation of few clusters that relatively soon coalesce, $\Delta \sim 0$, 
with a behavior quite in agreement with that suggested
by Eq.~(\ref{eq:simple}). 

However, simulations performed on larger systems 
to improve the clustering statistics revealed a rate of aggregation
different from that expected from Eq.~\eqref{eq:simple}
that basically applies only to binary collisions.
We, therefore, investigated
systematically the aggregation process through a set of
simulations on a system involving $N=10^4$ particles, where we
monitored the time evolution of the standard indicators used to
quantify the degree of clustering in aggregating systems
\cite{Torquato,Puglio}. Such indicators display a trivial
behavior when the noise level is $D > D_c$, as expected because 
uniform particle distributions there occur.


We measured the number of clusters $N_c(t)$,
their average mass $M_c(t)$, and the distribution $P_t(m)$
representing the number of clusters with size $m$ at
time $t$. We also considered the average distance $\Delta(t)$
between neighbor clusters which provides information about
the spatial compactness of the system. Such quantities were
averaged over a set of independent runs starting from random
particle distributions with a given density.

\begin{figure}
\includegraphics[clip=true,keepaspectratio,width=8.0cm]
{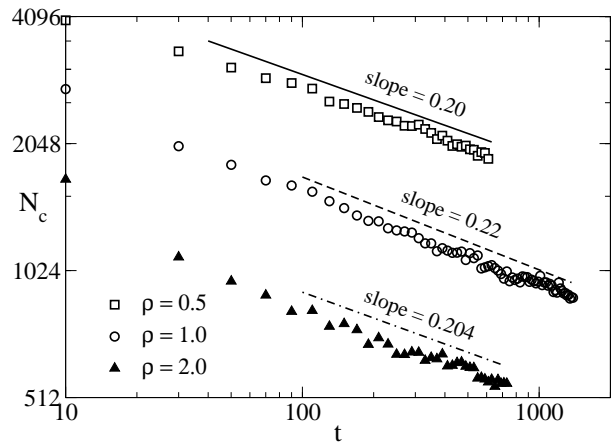}
\caption{Power law decay in time of the average number clusters $N_c$.
Data refer to noise $D = 0.05$ and to the 
three densities indicated in the inset.
Lines indicate the best fit to data whose slopes are the exponent 
$z$. The resolution is  $\varepsilon=0.1$ in all cases.}
\label{fig:Nclust}
\end{figure}

In all the cases considered we observed (see Fig. \ref{fig:Nclust}) the 
algebraic decay of the cluster number $N_c$ according to
\begin{equation}
N_c(t) \sim t^{-z}\;.
\label{eq:NCpow}
\end{equation}
Each cluster has a characteristic size and a typical distribution of 
particles that depend on the noise and the 
other parameters in agreement with the analytical result for the 
stationary single cluster distribution derived in Ref.~\cite{CeccoPRL}.   
This suggests that, when sufficiently isolated, a cluster 
persists over a longer period in a sort of quasi-equilibrium state.

The results from Fig.~\ref{fig:Nclust}, referring to systems with different
densities $\rho$ but same noise $D=0.05$, suggest an exponent 
$z$ close to $0.2$, independent of $\rho$ within
statistical errors. We also observed systematically that, at each time
$t$, the larger the density the lower the number of clusters.
This means that systems with higher 
densities generate a smaller number of clusters 
that, due to the conservation of the
number of particles, must contain a larger number of particles 
already in the first stages of their formation. 
The density therefore sensibly affects 
only the initial rate of cluster formation, that is, 
at higher densities clusters form faster. Later on, 
different densities reflect only in different
prefactors of the power law (\ref{eq:NCpow}). 
This also implies that, when working at too high densities, 
cluster statistics becomes worse and 
the power-law equation~(\ref{eq:NCpow}) cannot be clearly detected. 

\begin{figure}
\includegraphics[clip=true,keepaspectratio,width=8.0cm]
{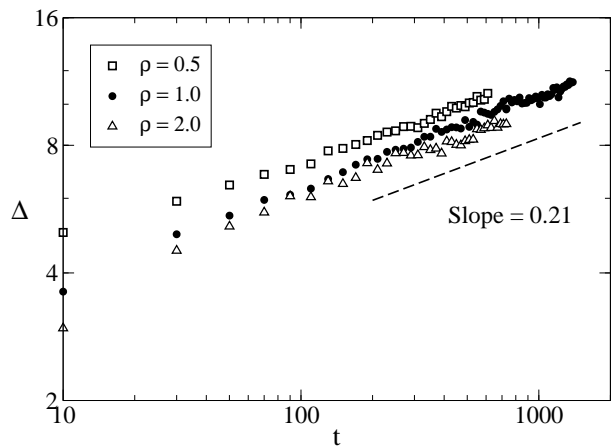}
\caption{Time-behavior of the average distance between clusters
for different particle density, for a system with noise $D = 0.05$.  
The resolution is  
$\varepsilon=0.1$ in all cases, however, the same data processed 
with resolution $\varepsilon = 0.5$ gave the same exponent within 
statistical errors.}
\label{fig:distance}
\end{figure}

\begin{figure}
\includegraphics[clip=true,keepaspectratio,width=8.0cm]
{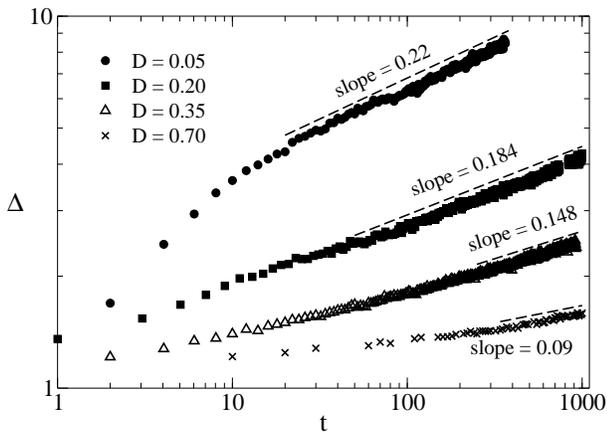} 
\caption{Power-law behavior with time of the average
cluster distance for a system with $N=10^4$ particles and different
noise levels $D$ at the same density $\rho=1.0$ and resolution 
$\varepsilon=0.1$.} 
\label{fig:delta_noise}
\end{figure}

A straightforward consequence of the conservation of the particle number and of
the decay (\ref{eq:NCpow}) is that the average cluster mass
$M_c$ grows with the inverse power law $M_c \sim t^{z}$, as 
verified in our simulations (not shown for the sake of space).
In figure \ref{fig:distance}, we plot the average distance
between nearest neighbor clusters $\Delta$ directly extracted
from the runs. Obviously, this quantity also exhibits the scaling
\begin{equation}
\Delta(t)  \sim t^{z}
\label{eq:distance}
\end{equation}
with the same exponent $z$ of the mean
cluster mass. The presence of such a power law seems to indicate
that as the process keeps going on and a reasonable number of larger
clusters is formed, further aggregation is suppressed and cluster
growth becomes very slow.
We also checked the dependence of $z$ on the noise strength
$D$.  Fluctuations induced by the high noise, indeed, produce 
fragmentation or evaporation of particles from already formed clusters, 
making the influence of $D$ on $z$ more relevant than that of
$\rho$. 
Figure~\ref{fig:delta_noise} reports data on the variation of the 
slope [Eq.~(\ref{eq:distance})] at different $D$ and indicates that,
although the power law behavior remains robust with respect to
$D$, until $D<D_c$, the $z$ exponent lowers as $D$ approaches from
below the critical value $D_c$.
Therefore, in the region of parameter we explored, the simulation
results indicate that the value of the exponent $z$  
weakly depends on the density but it is quite sensitive to the
noise level $D$.

\begin{figure}
\includegraphics[clip=true,keepaspectratio,width=8.0cm]
{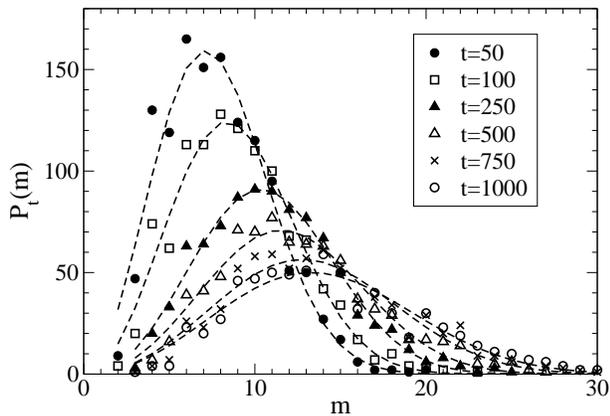}
\caption{Time-dependent histograms of the cluster masses collected
from simulations of a system of $N=10^4$ particles,
density $\rho = 1.0$, and $D=0.05$. Lines are guides for the eye.}
\label{fig:P_of_m}
\end{figure}

\begin{figure}
\includegraphics[clip=true,keepaspectratio,width=8.0cm]
{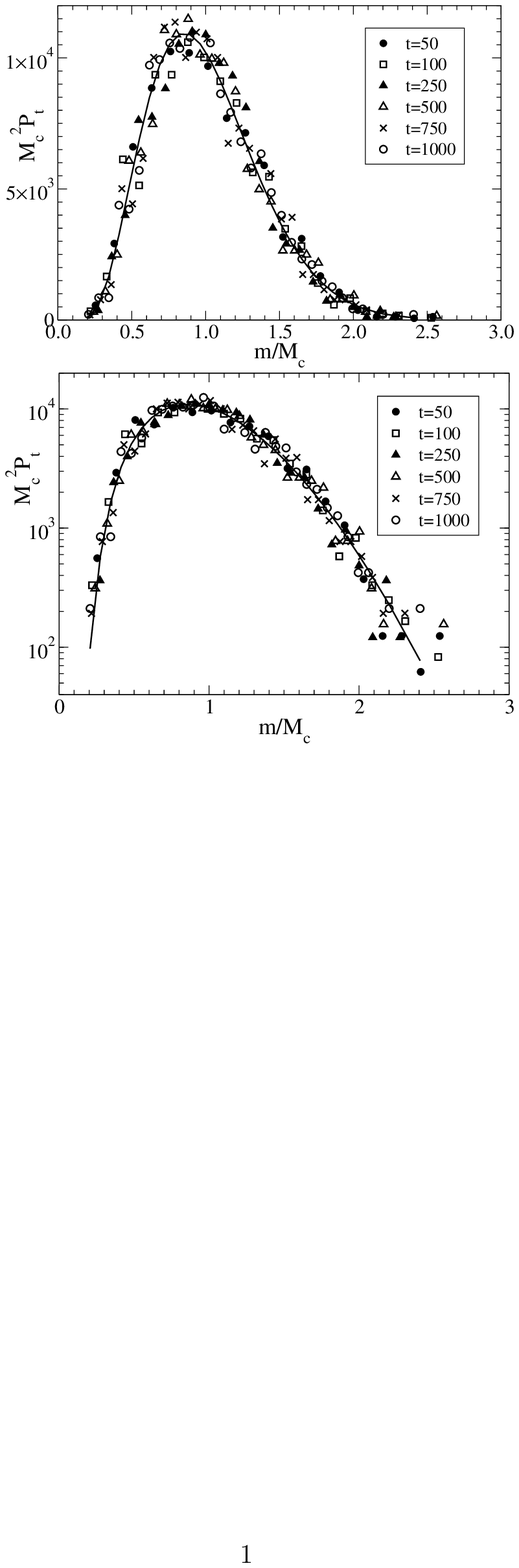}
\caption{Data collapse of the time dependent distributions shown in
Fig.~\ref{fig:P_of_m} obtained through the rescaling suggested by
Eq.~\ref{eq:Pscale}: linear-linear (top) and linear-log (bottom) plots. 
The solid line represents the fitting curve 
$f(x) = a_0 x^{-2}\;\exp\{a_1/x-a_2 x^2\}$.} 
\label{fig:collapse}
\end{figure}

To investigate in more details the  aggregation process
we collected the histograms of the cluster masses at different
times during the runs. The evolution of mass distribution is shown
in figure \ref{fig:P_of_m} at density $\rho=1$ and $D=0.05$.
The shape of the distribution, at earlier times, rises sharply  
at small values of $m$, from zero to a maximum and then decays mildly to 
zero at larger masses. 
The distribution are skewed and peaked around the average and their 
tails are not very extended. 
As time goes on,  
the average increases and the distributions become broader with
tails that spread over larger interval of masses $m$,  indicating 
a clustering activity that involves polydispersity in cluster size. 
Basic theories for aggregation phenomena \cite{General} 
suggest the scaling symmetry 
\begin{equation}
P_t(m)  \sim  M_c^\theta f(m/M_c)
\label{eq:Pscale}
\end{equation}
for cluster mass distributions for processes occurring 
with an asymptotically well-defined average size $M_c$ for the aggregates.
The constraint of the total mass (number of particle)
conservation, 
$$
N = \sum_{m} m P_t(m),
$$
sets the scaling exponent to the value $\theta = -2$ 
\cite{Redner,Leyvraz,Dongen}.
The rescaling of distributions in Fig.~\ref{fig:P_of_m}, according
to  Eq.~\eqref{eq:Pscale}, yields the data collapse  
shown Fig.~\ref{fig:collapse} in linear (top) and linear-log (bottom) 
plots.  


The collapse is rather satisfactory both for the bulk and tails of the 
distributions and fully consistent with the power-law
behavior of $N_c(t)$. In fact, 
as an immediate consequence of the scaling form~(\ref{eq:Pscale}), 
it is straightforward to verify that, in the continuum limit and with the 
change 
of variable $m \to m/M_c$, we obtain
$$
N_c(t) = \sum_{m} P_t(m) \sim M_c(t)^{-1}\;,
$$
{\em i.e.}, the number of clusters is proportional  
to the inverse of their average mass $M_c$.   
This analysis shows that our model, although not directly related
to general stochastic aggregation processes \cite{Smolu}, follows their  
typical scaling behavior.     

\section{conclusions}
In this paper we considered a one-dimensional model 
that promotes aggregation of particles for a certain range of the 
parameters. The model,  
related to a certain class of problems concerning the onset 
of social behavior among individuals, is defined by a first order 
stochastic differential equation. More specifically, 
each Brownian particle, representing a living organism, undergoes 
a drift velocity depending nonlocally on the other particle 
positions. This drift is the response to  
the population unbalance perceived by a single individual 
between left and right. 
Local fluctuations occurring in the population 
density determine a displacement towards the regions with the 
highest crowding. However,  
a limited sensing range of the individuals, mimicked in the model by 
an exponential weight in the distance, prevents them from reacting
to fluctuation events occurring too far.   
The competition between far large and near, but small,
aggregating centers makes the system dynamics nontrivial.   
Below a certain randomness threshold $D_c = 1$, 
aggregation  of particles in clusters of increasing size is observed, 
but at higher noise the organisms disperse via   
almost independent random walks suppressing social behavior. 

The aggregation process has been investigated through a set
of simulations where we monitored the time behavior of the
standard indicators used to quantify the degree of clustering in
aggregating systems \cite{Puglio,Torquato}. 
Such indicators of
course display a trivial behavior when the noise 
exceeds the threshold $D_c$  
because uniform particle distributions occur.
On the contrary they exhibit a power-law 
behavior at low noise regimes showing that this model obeys the typical  
scaling behavior of stochastic aggregation processes. 
However the relevant role of the spatial fluctuations, as already noted 
for other processes Ref.~\cite{Charge}, does not allow one to 
explain the exponents through the mean field theory defined by the  
Smoluchowsky equation for coagulation \cite{Smolu}. 


\begin{acknowledgments}
G.G. acknowledges the support by MIUR (PRIN 2004). 
G.P.S. acknowledges the support of CONICET.  
We are indebted to 
Amos Maritan, Matteo Marsili, Massimo Cencini and Paolo da Pelo 
for useful discussions and suggestions. 
\end{acknowledgments}

\appendix*\section{A}
In this appendix we show how to derive the  
evolution equation~(\ref{eq:simple}) for the centers of mass $X_1$ 
and $X_2$ of two aggregates, each containing half of the particles $N/2$.
We assume that this situation roughly corresponds to a particle distribution
with density 
$$
\rho(y,t) = \rho_0 \left\{\Theta[\delta^2/4 - (y-X_1)^2] + 
            \Theta[\delta^2/4 - (y-X_2)^2] \right\}
$$
where $\rho_0 = N/(2\delta)$, $\Theta(u)$ indicates the unitary step function
and $\delta$ represents the spatial extension of the two 
aggregates. Furthermore, we suppose that the intra-cluster dynamics is 
basically 
frozen. Therefore, each cluster travels as a rigid body maintaining its shape 
basically unchanged, so that the relevant time dependence lies only in their 
center mass positions $X_1(t)$ and $X_2(t)$.
This density produces a drift velocity that can be derived from the 
integral \eqref{eq:phi}
$$
\phi(X_1) =  \int_{X_1-\delta/2}^{X_1+\delta/2} dy \mbox{e}^{-(y-X_1)} 
            + \int_{X_2-\delta/2}^{X_2+\delta/2} dy \mbox{e}^{-(y-X_1)}\;,  
$$
then
$$
\phi(X_1) = 2(1-\mbox{e}^{-\delta/2}) + 
2\sinh(\delta/2)\mbox{e}^{X_1 -X_2}\;.
$$  
According to Eq.~\eqref{eq:langev}, 
the velocity of cluster $1$ reads 
$$
V_1 = \frac{\phi'(X_1)}{\phi(X_1)} = 
\frac{\sinh(\delta/2)\mbox{e}^{X_1 -X_2}}{(1-\mbox{e}^{-\delta/2}) 
+ \sinh(\delta/2)\mbox{e}^{X_1 -X_2}} \;,
$$
that, after some algebraic manipulations simplifies to 
\begin{equation}
V_1 = \frac{1}{1+ b\mbox{e}^{X_2-X_1}} 
\label{eq:V1}
\end{equation}
with $b = [1-\exp(\delta/2)]/\sinh(\delta/2) = 2/[1+\exp(\delta/2)]$. 
The velocity of the second cluster,  
due to the symmetry of the problem, 
has to be 
$V_2 = -V_1$, so that the relative velocity of the two rigid clusters 
is $-2 V_1$, that is Eq.~(\ref{eq:simple}) providing that 
$\Delta = X_2-X_1$. 
More generally, we can observe that for two very localized aggregates 
of sizes $\delta_1$ and $\delta_2$ with local density $\rho_1$ and  
$\rho_2$, respectively, the integrals contributing to $\phi$ can be 
roughly approximated to $\rho \delta$, the  
cluster relative velocity is thus estimated as 
$$
V_r = 2\frac{\rho_2\delta_2\mbox{e}^{-\Delta}}
{\rho_1\delta_1 + \rho_2\delta_2 \mbox{e}^{-\Delta}} 
= \frac{2}
{1 + b \mbox{e}^{\Delta}}\;.   
$$ 
The parameter $b = \rho_1\delta_1/(\rho_2\delta_2)$ thus 
represents a measure of the relative weights of the two clusters.

\end{document}